\documentclass[apj]{emulateapj}
\setkeys{Gin}{width=0.48\textwidth}
\usepackage{graphicx}

\newcommand{\kpc}{{\rm kpc}}

\newcommand{\hmpc}{\ifmmode{h^{-1}\,\hbox{Mpc}}\else{$h^{-1}$\thinspace Mpc}\fi}

\newcommand{\kms}{\ifmmode{\,\hbox{km\,s}^{-1}}\else {\rm\,km\,s$^{-1}$}\fi}
\newcommand{\msun}{{\rm\,M_\odot}}

\begin{document}
\title{Gaps in the GD-1 Star Stream} %\altaffilmark{1}}
\shorttitle{GD-1 Gaps}
\shortauthors{Carlberg \& Grillmair}
\author{R. G. Carlberg}
\affil{Department of Astronomy and Astrophysics, University of Toronto, Toronto, ON M5S 3H4, Canada} \email{carlberg@astro.utoronto.ca }
\author{C. J. Grillmair }
\affil{Spitzer Science Center, 1200 E. California Blvd., Pasadena, CA 91125, USA}
\email{carl@ipac.caltech.edu}
%\altaffiltext{1}{Department of Astronomy and Astrophysics, University of Toronto, Toronto, ON M5S 3H4, Canada carlberg@astro.utoronto.ca }
%\altaffiltext{2}{Spitzer Science Center, 1200 E. California Blvd., Pasadena, CA 91125, USA, carl@ipac.caltech.edu}

\begin{abstract}
GD-1 is a long, thin, Milky Way star stream that has readily visible density variations along its length. 
We quantify the locations, sizes and statistical significance of the density structure,  {\it i.e.} gaps, using a set of scaled  filters.
The shapes of the filters are based on the gaps that develop in simulations of dark matter sub-halos crossing a star stream. 
 The high Galactic latitude 
8.4 kpc long segment of GD-1 that we examine has $8\pm3$ gaps of 99\% significance or greater, with the error estimated on the basis of tests of the gap-filtering technique. 
The cumulative distribution of gaps more than 3 times the width of the stream is in 
good agreement with predictions for dark matter sub-halo encounters with cold star streams.
The number of gaps narrower than 3 times the width of the GD-1 stream 
falls well below the cold stream prediction which is taken into account for the gap creation rate integrated over all sizes. Simple warm stream simulations scaled to GD-1 show that the falloff in gaps 
is expected for sub-halos below a mass of $10^6 \msun$.  
The GD-1 gaps requires 100  sub-halos $>10^6\msun$within 30 kpc, the apocenter of GD-1 orbit. 
These results are consistent with LCDM sub-halo predictions but further improvements in stream signal-to-noise and gap modeling will be welcome.
\end{abstract}

\section{INTRODUCTION}
\nobreak

\begin{figure*}
\begin{center}
\includegraphics[angle=0,bb= 44 336 572 454,scale=2.1]{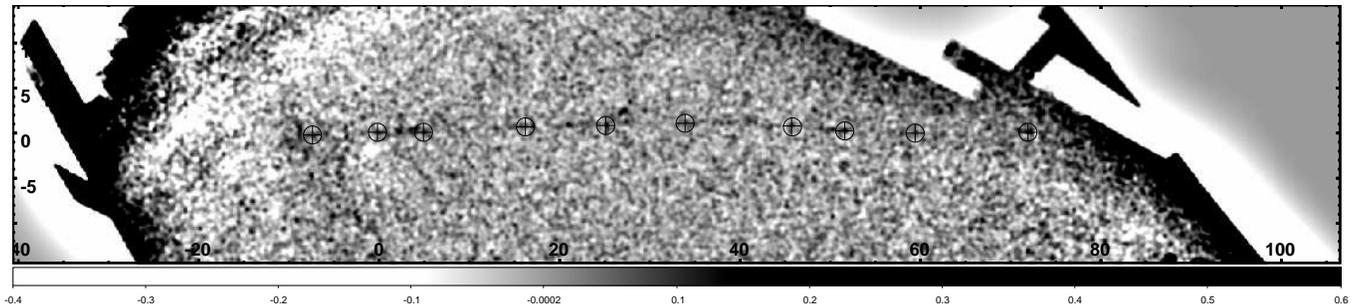}
\end{center}
\caption{An image of the stream in a rotated sky coordinate system in which the stream is placed near the new equator. The coordinates are in degrees  relative to the second cross-hair on the figure and are not aligned with any standard system. To remove the varying background, the masked image has been smoothed over 5\degr, subtracted from the original image, and then smoothed with a 0.4\degr\ Gaussian.  The ``cross-hairs'' are used as the initial guess of the centerline used to extract the stream.}
\label{fig_image}
\end{figure*}

The GD-1 star stream was discovered in the Sloan Digital Sky Survey (SDSS) Data Release 4 photometry \citep{GD:06} with 
Data Release 7 improving both the sky coverage and photometric uniformity \citep{Willett:09,KRH:10}.  
The stream arcs more than 80\degr\ across the northern sky, passing within 30\degr\ of the Galactic pole. 
The visible section of the stream is about 8 kpc distant at its midpoint with the distance increasing
 from about 7 to 11 kpc.
The stream is exceptionally narrow, having an angular width of  0.5\degr,  which corresponds to a linear width of about 70 pc. The stream has a remarkable length to width ratio of at least 100. Figure~\ref{fig_image} rotates the sky to place the stream near the reoriented equator in a Mercator projection.

SEGUE spectra \citep{Yanny:09} provide velocities and metallicity measurements of the stars in selected regions along the stream. 
Those  velocities, augmented with proper motions \citep{Munn:04,Munn:08} from the comparison of the SDSS to the USNO catalogs \citep{Willett:09,KRH:10},
provide sufficiently accurate phase space information along the stream
 to give a good GD-1 orbit and some limits on the shape of the Galactic potential. 
The derived orbits have  a perigalacticon of about 14 kpc and apogalacticons of  26-29 kpc.
The orbits passes through the plane of the galaxy at large radii where there are
 relatively few high density molecular gas clouds or HII regions.
The narrowness of the stream and the stellar mass in the stream, $2\times 10^4 \msun$ \citep{KRH:10}, suggest that the likely progenitor is (or was) a globular cluster \citep{GD:06}.  
No progenitor for the GD-1 stream has yet been identified close to the orbit of the stream.

The stream has readily visible density variations along its length, which 
are the primary interest of this paper. 
\citet{KRH:10} simply stated that the reason for the density variations is not clear.  
For our purposes,  an absent or distant progenitor is a useful simplification 
in that the epicyclic density variations discussed in \citet{Kupper:08,Kupper:12} 
play no role in the density variations of the stream. 

The GD-1 stream allows a fairly strong test for the existence and properties of the  large population of sub-halos, $N(>M) \simeq 1.6\times 10^5 (M/10^6 \msun)^{-0.9}$, 
orbiting within a Milky Way-like dark halo (counted to a distance of 433 kpc) 
predicted to exist in dark halos simulated from LCDM initial conditions \citep{VL1,Aquarius,Stadel:09}. 
When a sub-halo passes through a stellar stream it normally pulls out a loop of stars 
which develops into a visible gap in the stream density \citep{Carlberg:09,YJH:11}.  
The rate per unit length of gap creation, $R_\cup$,  is derived in the cold stream approximation as 
$R_\cup \propto M_m^{-0.36}$ \citep{Carlberg:12}, where $M_m$ is the smallest mass sub-halo that can create a visible gap of any size greater than the stream width.  

We previously developed a density filtering procedure to locate gaps and assign a 
statistical confidence level to their detection which we applied to 
the  Pal~5 stream \citep{CGH:12}. 
Now that improved GD-1 data are available 
it is a better case for the study of gaps. That is, 
Pal~5 is significantly more distant than GD-1, some 23 kpc as compared to 8 kpc, respectively, and, 
the Pal~5 stream skirts the Galactic plane in projection, hence has a fairly large stellar background. 
Furthermore, the Pal~5 globular cluster is very much present with the galactic tides still pulling stars away 
from the cluster on orbits that are expected to create density variations in the stream near to the cluster \citep{DOGR:04,Kupper:12}. 
The GD-1 star stream is a high galactic latitude system, nearly a segment of a great circle  when viewed in re-oriented coordinates (see Figure~\ref{fig_image}) \citep{Willett:09,KRH:10}. 
It has a relatively high signal-to-noise along its length and is one of the best streams to search for gaps and to test theories for their evolution in the galactic potential.

In this paper, we first search for the gaps along the GD-1 stream using a set of filters based on the density profiles of gaps that develop in simulations of sub-halo stream interactions.  
We characterize the statistical behavior of the filters on a model stream with gaps inserted.
The numbers of gaps in the GD-1 stream and the distribution of their sizes are
compared the predictions from  dark matter sub-halos to estimate the 
lowest effective sub-halo mass and their numbers 
within the orbit of the stream.

\section{THE STREAM MAP AND DENSITY PROFILE}

The SDSS Data Release 8 \citep{SDSS,DR8} extinction corrected stellar photometry is optimally  match-filtered in
 color-magnitude space 
to identify old metal poor stars at distances around 8 kpc.
The filtering uses the procedures documented in 
\citet{Rockosi:02,GD:06,Grillmair:09} and \citet{Grillmair:11}. 
The pixel values are star counts filtered on the basis of their agreement with the M13 color-magnitude relations 
and weighted with a luminosity function for a $Z=0.0003$ low metallicity population. 

The distance to the GD-1 stream varies along its length. 
The same color-magnitude  filter has been offset +0.2 magnitudes to make four separate sky maps
with approximate distances ranging from 6.7 kpc for the first plane to 8.8 kpc for the last 
(based on an assumed distance of 7.7 kpc for M13, the basis of the filter). We will present the density profiles of all four maps below.
Most of the results below use the 8.1 kpc distance version (the third of the four maps) on the basis that it has the highest total stream density. 

\begin{figure*}
\begin{center}
\includegraphics[angle=0, bb= 48 354 566 430, scale=2]{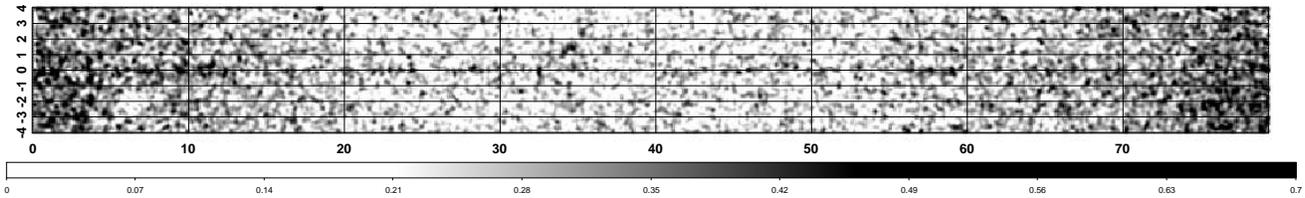}
\end{center}
\caption{An image of the extracted stream relative to the optimized centerline.   
The image is smoothed with a 0.2\degr\ Gaussian.
 The coordinates are in degrees relative to the centerline
with x and y and coordinates being along and transverse to the stream, respectively. No compensation for the varying background has been applied. Our analysis is applied to the  [10,70] degree range of this image.}
\label{fig_xy}
\end{figure*}

We rotate the map from its J2000 Mercator coordinates to a new Mercator projection 
with the stream along the new equator. 
The (arbitrary chosen) stream point originally at [RA, Dec] = [141.80\degr, 27.04\degr], where the second cross-hair symbol is located, becomes the new [0,0] 
and the new equator passes through the point [RA, Dec] =[222.29\degr, 58.22\degr].   
The rotated image is placed into the same size pixels, 0.1\degr, as the original image, which introduces correlations between 
adjacent pixels, but the pixels remain statistically independent at larger separations.  
Figure~\ref{fig_image} shows the image in the new coordinate frame. 
For better viewing a 5\degr\ Gaussian smoothed image 
is subtracted to approximately remove the variation in the stellar density over the sky.
The circled plus signs show the points that serve as our initial estimates of the centerline of the stream.  
Although the path of the stream can be identified further after it crosses the Sagittarius stream (the broad feature that runs through [-5,0] on the left side of the map)
we do not include that portion in our analysis since it has undesirably large errors from the 
large and variable background correction. 
On the right side it is also unclear exactly where the stream goes, 
although we do note that our 70\degr\ point extends the stream a few degrees further than  previous papers.

The segment of the GD-1 stream that we have selected is above 40\degr\ galactic latitude, 
where the E(B-V) values from \citet{SFD:98} are generally below 0.02 mag.  
Any extinction corrections that equally affect the stream and background region would make 
little difference to the background subtracted stream density.  
Because the total extinction corrections are so low, 5\% or less, in a stream of about 20\% over-density, 
any errors that may exist in the extinction corrections are not sufficient
to cause gap-like density variations.

The stream is given xy coordinates relative to the centerline.
The initial placement of a set of centerline points is done by hand on local density peaks, after which a simple random search is undertaken to maximize the mean stream density over the interval [10,70].
The optimization procedure moves the points less than a pixel on average and boosts the density relative to the initial guess by about 10\%. A spline interpolates the centerline between the points.
The x coordinate is measured as distance along the fitted centerline, relative to the first point which is at the edge of the 
Sagittarius stream.
The  y coordinate is the distance to the interpolating spline.  
The centerline is a small amplitude, slowly varying curve. However there are some small scale local excursions
which are not removed.  

Perhaps the most interesting local excursion is located near  x = 50\degr\ in Figure~\ref{fig_xy}, 
where the stream has an s-like shape above and below the axis. 
The feature is visible in Figure~\ref{fig_image} around  42\degr of its rotated coordinate system (which is not the same as 
the Figure~\ref{fig_xy} system).
The feature, although not very compelling at the available signal-to-noise, 
could be the result of an encounter with a sub-halo
 or possibly be the location of a recently disrupted progenitor object.
We consider an alternate centerline through these points below. 

We estimate the  background using the regions on both sides of the stream beyond $\pm$0.5\degr\ where no stream signal is visible.
The background region extends 5 times the width of the selection region, 
that is, from 0.5 to 1.75\degr\ distant from the centerline on either side of the stream 
in the image with no smoothed background subtracted. 
This background is subtracted from the total density in the centerline region to yield the linear density of the stream.
The background subtracted density is extracted into bins of 0.1\degr\ along the stream which we use in the analysis of the GD-1 stream properties.

\begin{figure}
\begin{center}
\includegraphics[scale=0.6]{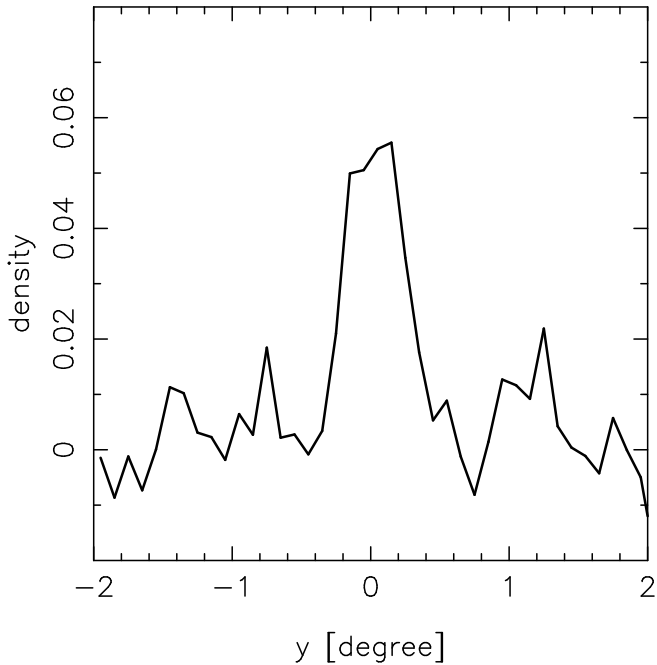}
\end{center}
\caption{The density profile across the stream, averaged from x = 10-70\degr. 
The FWHM is 0.50\degr. 
The stream is extracted within $\pm0.25\degr$ of the fitted centerline and the background region starts at $\pm0.5\degr$.}
\label{fig_width}
\end{figure}

The stream densities are summed along its length from 10-70\degr\ in Figure~\ref{fig_xy} to create a transverse density profile.
We  measure the FWHM of the stream (simply using the full width at half the peak height)
over the 60 degree length to be 0.50\degr, as shown in Figure~\ref{fig_width}.  
\citet{KRH:10} over a similar length of the stream found a Gaussian width of 0.20\degr, which, 
for a Gaussian, converts to a FWHM of 0.47\degr, which is consistent with our measurement. 
To create a density profile along the length of the stream we sum the density over $\pm$ 0.25\degr\ around the centerline which captures about 90\% of the stream stars.
 Larger stream widths slightly increase the total density of the stream, but it leads to a decrease of the signal to noise.

\begin{figure}
\begin{center}
\includegraphics[angle=-90, scale=0.7]{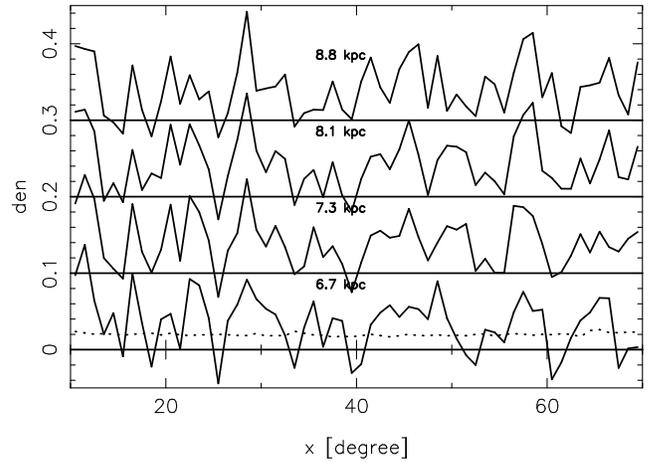}
\end{center}
\caption{The background subtracted density along the stream in 1\degr\ bins with distance increasing from the bottom upwards, subsequent lines offset by 0.1 in density. The dashed line indicates the estimated error in each bin.}
\label{fig_nx4}
\end{figure}

An estimate of the point by point error of the measured stream density is calculated from the dispersion in the estimate 
of the local background density, $\sigma_i$. Since the background is $N$ times the width of stream region the background noise in the stream region will be $\sqrt{N}\sigma_i$. The mean density in the stream is a factor $\delta$ higher than the background which boosts the error in the stream to  $\sigma_i\sqrt{N(1+\delta)}$. Subtracting the background from the stream adds its error of $\sigma_i$ in quadrature to the stream error to give an error for the background subtracted stream to be $\sigma_i\sqrt{N(1+\delta)+1}$. 
The GD-1 stream has $\delta$=0.178 in our 8.1 kpc map and normally we use a background width 
such that $N=5$ although we do  the analysis for $N=7$ as well to check for anomalies. 
Including the stream over-density in the variance calculation boosts the required peak height for 99\% confidence by about 5\%.

The density profiles for all four distances of the matched filter are displayed in Figure~\ref{fig_nx4}. Each has had its own search for an optimal centerline and each uses its own background calculation. The profiles are highly correlated.
The mean density in the streams is, from nearest to farthest, 
0.147, 0.170, 0.178, and 0.166, 
respectively. Given the relatively minor differences between the profiles 
we prefer to use the profile with the largest mean density, 
the one at a mean distance of 8kpc  (the second from the top in Fig.~\ref{fig_nx4}) as the best to search for gaps.
However, we do analyze all four distance maps below.

\section{STREAM GAP-FINDING}

The passage of a dark matter sub-halo, or in principle any massive object, through or close to the stream causes 
material in the region of encounter to gain or lose angular momentum and spread apart creating a gap in the middle and density pileups on either side \citep{Carlberg:12}. 
To find such gaps we use the density profile filtering approach  initially developed in \citet{CGH:12}, augmented below
with a study of its statistical performance for the GD-1 stream parameters.

\subsection{Scaled Gap-Filters}

The first step in gap-finding is to convolve the stream density profile
with a gap filter which approximates the shape of the gaps found in simulations. 
We use gap-filters with the functional forms $ w_1(x)=(x^6 -1)\exp{(-1.2321 x^2)}$ and
$ w_2(x)=(x^8 -1)\exp{(-0.559 x^4)}$ both of 
which have a low density floor and compensating peaks on either side so that the filters have zero mean over the range $x=[-3,3]$.  
The $w_2$ filter has a quicker descent to the floor and sharper compensating peaks than $w_1$.  
The filter is scaled to gap lengths of 0.1\degr\ to 5\degr\ along the stream.

The second step in gap-finding is to establish the confidence levels of the peaks in the filtered density profile, 
that is, the probability that a peak in the filtered density field is a gap detection and is not a peak that results from the noise in the stream density measurement. 
A confidence level of, say, 99\%, means that there is only a 1\% chance that the peak is the result of chance fluctuations in a noisy but otherwise gap-less density profile.

We use the measured background density profile as the starting point to construct random density profiles
 which do not contain any gaps. 
The background density field is first flattened with a third order polynomial  to remove any gradients along the stream density profile which can artificially increase the scatter around the mean stream density.
The flattened background density field is then scaled to have the same estimated noise level as the stream. 
This gap-free density field is then sampled, with replacement, 
to create 100,000 random density profiles which are then analyzed for peaks in precisely the same way as the stream. 
It is this step that requires that the background density profile have an approximately constant mean density, otherwise in the resampling points from regions of different mean densities are mixed which leads to an increased number of peaks being found. 
The gaps found in the random realizations are sorted to determine the filtered field heights required to be above
some required confidence level. We normally use the 99\% level in this paper.
After applying filters of all lengths to the stream density profile we select the most significant peak 
 above 67\% probability present at any location in the stream. 
This procedure means that lower significance narrow gaps are appropriately incorporated into higher significance wider gaps. 
The edges of the highest significance peak detection at each location, as defined by the filter scaling, is used for the purpose of illustration to create a square shouldered gap with a depth of 0.5 divided by its confidence level relative to the required level as shown in Figures~\ref{fig_demo} and \ref{fig_gaps}.

\subsection{Gap Filter Tests}

\begin{figure}
\begin{center}
\includegraphics[angle=-90, scale=0.7]{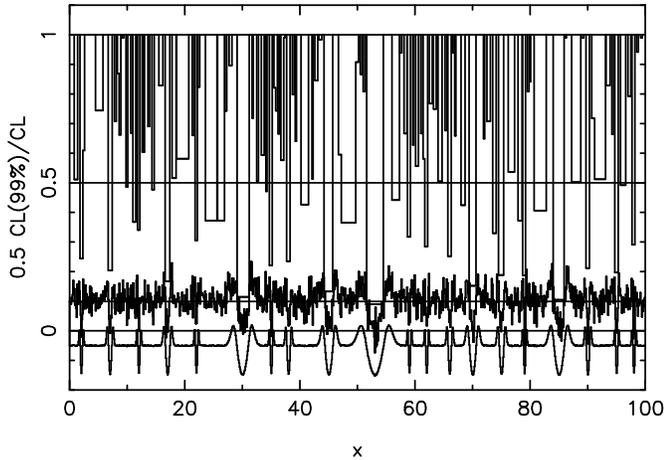}
\end{center}
\caption{A demonstration of the gap filtering procedure with test data. 
The noiseless stream is shown at the bottom and the same stream with added noise is above it. 
The 99\% confidence gaps that the filtering procedure finds are the square profiles that are below a value of 0.5 in the plot.}
\label{fig_demo}
\end{figure}

We undertake a number of simple tests of the statistical reliability of the gap-finding procedure
for a stream with the noise characteristics of GD-1.
Non-overlapping  gaps are inserted into a uniform density using our filters (normally  $w_1$ is used) 
as shown in the lower line in Figure~\ref{fig_demo} to create an artificial stream with gaps.
Gaussian noise is added at each point as shown as the noisy density profile in 
Figure~\ref{fig_demo}. 
The measured RMS fluctuations in the data are 34\% of the mean level, but that also includes the extra fluctuations from whatever gaps are present. 
The artificial stream has noise added with $\sigma=0.3$ around a mean of one.
We also create a realization of the background with the same noise level.  
These two profiles are then analyzed in exactly the same way as real stream data. 
The resulting gap detections are shown in Figure~\ref{fig_demo}.  

\begin{figure}
\begin{center}
\includegraphics[angle=0, scale=0.8]{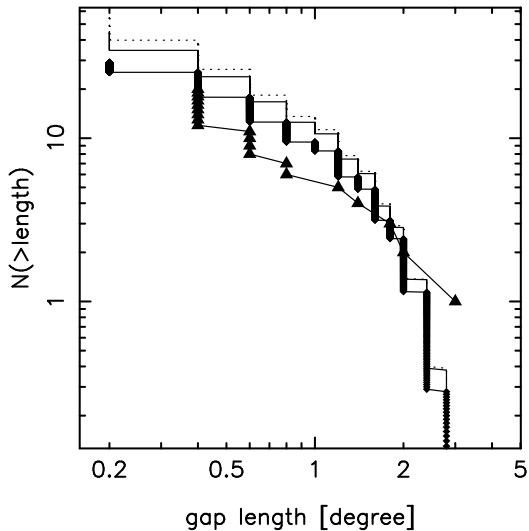}
\end{center}
\caption{Comparison of the distribution of gap sizes input (triangles), and the mean of 100 random distributions at 90 (dotted) 95  (solid line) and 99\% confidence (line with small diamonds) found for $\sigma=0.3$  with the $w_1$ filter.  }
\label{fig_manygaps}
\end{figure}

The statistical differences between the distribution of input gap sizes and the gap-filter results is quantified in Figure~\ref{fig_manygaps} which shows the input distribution of gap sizes (triangles) and the recovered 95 and 99\% confidence distributions for the $w_1$ filter. 
The $w_2$ filter is not well matched to the inserted $w_1$ gaps 
and always has more false positives than the $w_1$ filter so we do not consider it further. 
The $w_1$ filter performs well for 99\% confidence gaps even at a noise level enhanced above what we measure in the data.
The gap distributions from the $w_1$ filter at $\sigma=0.3$ noise for 100  realizations of the added noise are shown in Figure~\ref{fig_manygaps}.  
On average the total number of recovered gaps at 99\% confidence level is $1.44\pm0.15$ times the input number.  
We will take the mean bias to be a factor of 1.44 with a 3 standard deviation ($\simeq$99\% confidence) spread of 30\%.
An extremely important outcome is that at small angles the
recovered distribution is shifted upwards but otherwise retains the same  basic $N(>\ell)$ shape with no induced features.

The biggest gaps are biased to small sizes, but  less than Figure~\ref{fig_manygaps}  suggests. 
That is, the procedure finds that the largest recovered gap is on the average $2.4\pm 0.4$ whereas the input value 
 is 3.
Consequently the recovered largest gaps is on the average  at 1.5 standard deviations low, but well within the population distribution. Furthermore we caution that this result is for only a single realization of a gap distribution. 
Further efforts to improve the performance of the gap-filtering procedure are clearly warranted.

These simple tests also underscore that the quality of the results will depend on the shapes of the gaps and the details of the noise distribution. We note that our assumed gap density profile and the Gaussian noise that we have added are only approximate representations of the properties of the real data.

%==> tmp.s3190 <== 0.2 58.1818
%==> tmp.s3195 <== 0.2 46.5455
%==> tmp.s3199 <== 0.2 30.5455
%[carlberg@eastend]# tail --lines=1 -q s3_*/tmp.199 | awk '{print $2}' | cmd.msd
%101 28.8218 2911 3.01283 0.301283
%[carlberg@eastend]# tail --lines=1 -q s3_*/tmp.195 | awk '{print $2}' | cmd.msd
%101 44.505 4495 4.58716 0.458716
%[carlberg@eastend]# tail --lines=1 -q s3_*/tmp.190 | awk '{print $2}' | cmd.msd
%101 56.4752 5704 5.28513 0.528513

%tmp.a below 99, 95, 90
%0.2 1.51435 2.17943 2.59808
%0.4 1.61755 2.12539 2.35737
%0.6 1.76555 2.2488 2.45933
%0.8 1.8042 2.28671 2.47552
%1.2 1.65657 2.0202 2.10101
%2.0 1 1.20454 1.20454
%3 0.136364 0.136364 0.136364

\subsection{Gaps in the GD-1 Stream}

\begin{figure}
\begin{center}
\includegraphics[angle=-90,scale=0.7]{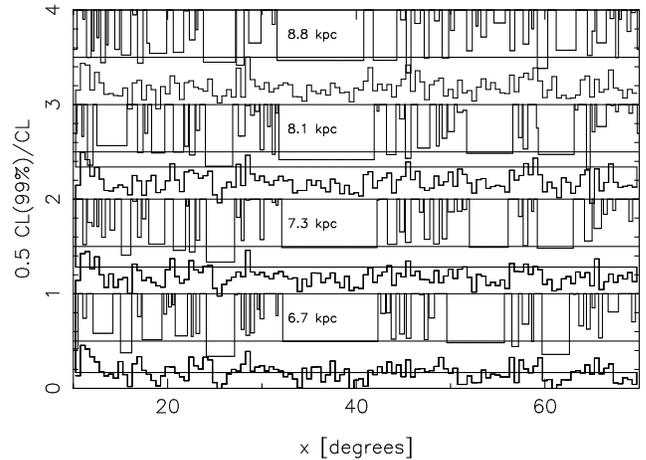}
\end{center}
\caption{The GD-1 reconstructed gaps at 99\% confidence with distance increasing from the bottom upwards. 
The density profiles in 0.5\degr\ bins are also shown. Our primary results are for the 8.1 kpc distance, second from the top.}
\label{fig_gaps}
\end{figure}

The results of gap-filtering the GD-1 stream with the $w_1$ filter are displayed in Figure~\ref{fig_gaps}. 
 The $w_1$ filter finds that there are 12 gaps above 99\% confidence. 
Our tests of gap-filtering indicates that we should reduce the 99\% confidence results by a factor of 1.44, to give 8.3 gaps and then assign a 30\% error to the result, so $8\pm3$ at 99\% confidence. Although we prefer our 99\% values it is interesting to note that at 95\% confidence the same procedures find 18 gaps in the GD-1 stream segment. The statistical tests find that this is a factor of 2.2 above the true number, which then becomes  8, the same as estimated at 99\% confidence. 

We undertake precisely the same gap analysis on each distance slice to check for anomalies and significant variations. From nearest to farthest there are 7, 6, 12 and 11  99\% gaps, respectively, with the third distance being our preferred choice. 
At 95\% confidence the numbers are 21, 20, 19, and 22 gaps, respectively, although our testing indicates that about half of these are false positives.  However, these results indicate that the streams at each distance are strongly correlated and as a consequence, so are the gaps that are found.

%==> SIT1/tmp.195 <==
%0.199999999999996 21
%==> SIT2/tmp.195 <==
%0.199999999999999 21
%==> SIT3/tmp.195 <==
%0.199999999999999 18
%==> SIT4/tmp.195 <==
%0.199999999999999 21

The region around 50\degr\ of Figure~\ref{fig_xy} contains an s-like feature around 50\degr\ which is not traced by our slowly varying streamline function. 
If we force the streamline to go through these points it increases the mean density in this region
and changes the details of the gaps found in the region. For either centerline of the stream,
the confidence level of the 3 to 4 gaps in the region is not quite 99\%, but above 95\% confidence.
The presence of gaps in this region suggests that it is unlikely to be young, {\it i.e.} recently released from a progenitor.
At the present time these small deviations from the smooth path of the stream are not very statistically significant and are not used in the analysis, however they are potentially a very powerful tool as the stream signal-to-noise is improved in future observations.

\section{GAP STATISTICS AND SUB-HALOS}

\subsection{The Stream Age}
The mean dynamical age of the stream, that is, the average time since the stream stars became unbound from the progenitor, is a key parameter for the interpretation of the stream gaps. 
The ages of the stars are estimated from the the photometry to be about 9 Gyr \citep{KRH:10}, 
which sets an upper limit to the stream age.  
A simple estimate of the dynamical age of the stream assumes that the stars started near a common point (which rotates with time) but with a small spread in angular momentum that causes the stars to spread apart along nearby orbits. 
As the low and high angular momentum stars drift apart, the width of the stream becomes entirely 
dependent on the dispersion of random velocities perpendicular to the stream orbits.
The radial velocity dispersion is approximately $\kappa a$, where $a$ is the typical epicyclic size and $\kappa$ is the epicyclic frequency. 
A locally flat rotation curve has $\kappa  = \sqrt{2} \Omega$,  the local circular velocity frequency.  Taking $a=0.035$ kpc at half the stream width, 
$\Omega= 220 \kms/ 15 \kpc $,  we find that the velocity dispersion is approximately 1 \kms, well below the 7 \kms\ upper limit measured in \citet{KRH:10}.
For a spreading rate of 1-2 \kms\  it will take  4.6-9.2 Gyr for stars to spread along the the 8.4 kpc long stream segment, if the only source of spreading was a range of angular momentum in the stream at some nominal time of unbinding from the progenitor. 
Assuming one end is relatively young then the mean age is 2.3-4.6 Gyr. 

\subsection{The Distribution of Gap Sizes}

\begin{figure}
\begin{center}
\includegraphics[scale=0.8]{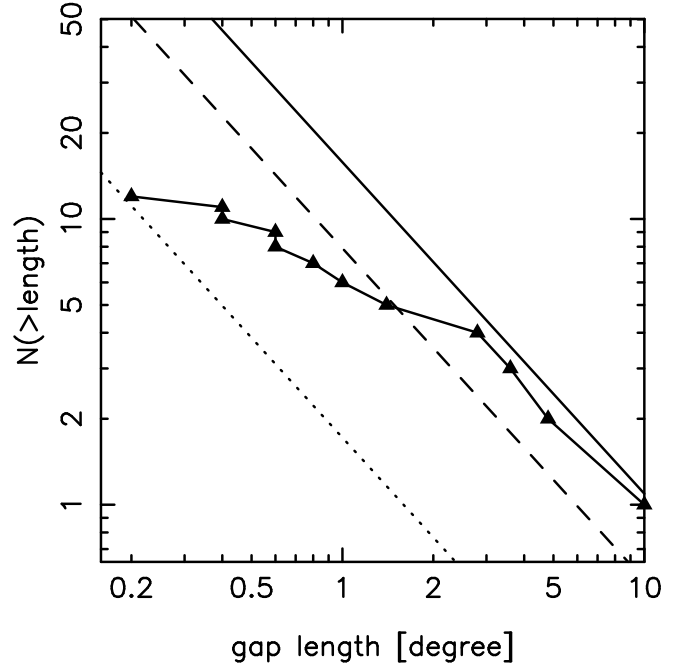}
\end{center}
\caption{The cumulative distribution of gap sizes at  99\% confidence confidence. 
The lines are estimates of the expected distribution, 
drawn for a mean age of 4.6 Gyr (solid), 2.3 Gyr (dashed) and 0.5 Gyr (dotted).}
\label{fig_ngap}
\end{figure}

The high confidence gaps in the GD-1 stream provides an opportunity to examine the distribution of gap widths. 
The cumulative distribution of the number of gaps wider than some size $\ell$, $N(>\ell)$, 
is displayed in Figure~\ref{fig_ngap} for 99\% significance gaps. 
The interesting feature of the plot is that the numbers do not continue to rise for gaps narrower than about 1-2\degr. 
The measured FWHM of the stream is 0.5\degr. equivalent to 70 pc. 
The \citet{Carlberg:12} prediction of the total number of gaps visible assumed 
that gaps as narrow as the width of the stream were readily visible whereas in  Figure~\ref{fig_ngap} the rise in the
numbers of gaps 
flattens for gap sizes smaller than approximately 3 times the 0.5\degr\ width of the stream.

The shallow s-curve near x = 50\degr\ of Figure~\ref{fig_xy} has gaps within it. 
Forcing the centerline through this feature does not significantly change the distribution of gaps,
 although it does slightly increase the amplitude and signficance of the features near 50\degr. 
If a stream progenitor disrupted at this location in the last 0.5 Gyr we would not expect any significant sub-halo induced gaps to have developed in the short time available. 
On the other hand, 
\citet{Kupper:08,Kupper:12} have established convincingly that a dissolving globular cluster will have a few epicyclic gaps visible nearby.
As a cluster dissolves its tidal radius, which is the characteristic radius of the epicycles, 
goes to zero so that the features shrink in proportion and should become effectively invisible  \citep{Kupper:10}.
Therefore a secondary conclusion is that the 50\degr\ feature, although not of high statistical significance in the first place, is unlikely to be the location of a recently dissolved progenitor.

\begin{figure}
\begin{center}
\includegraphics[angle=-90,scale=0.7]{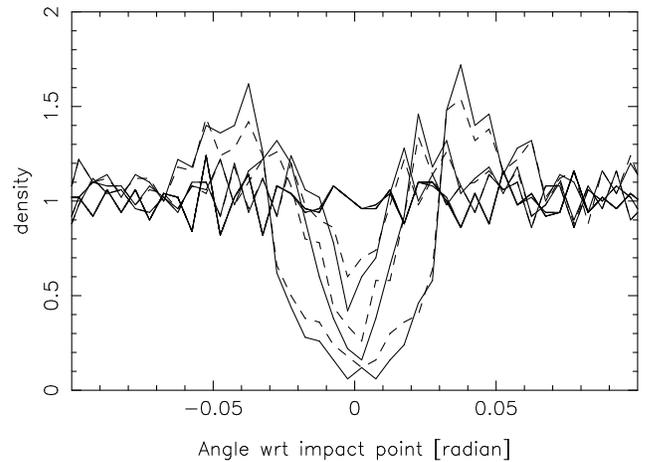}
\end{center}
\caption{The gap density profiles for a 70 pc wide stream encountering sub-halos at 90 (solid lines) and 150 pc (dashed lines) from the stream centerline. 
The three pairs of lines of increasing width and depth are for masses
$3\times10^5$, $10^6$ and $3\times 10^6\msun$, respectively, for the two distances of closest approach to the centerline. 
The lines across the center are the initial density distribution and indicate the noise level.
}
\label{fig_gapsM}
\end{figure}

The cold-stream analysis of \citet{Carlberg:12}, refined for galactic radii less than 30 kpc, 
predicts a gap creation rate as a function of galactocentric distance, $r$, in units of 30 kpc, and $M_8=M/10^8\msun$ of,
\begin{equation}
R_\cup =0.00433r^{0.26} M_8^{-0.36} {\rm kpc}^{-1} {\rm Gyr}^{-1}.
\label{eq_R}
\end{equation}
Sub-halos  of mass $M_8$ create gaps of mean length $\ell$,
\begin{equation}
\ell = 9.57 r^{0.16} M_8^{0.31} {\rm kpc}.
\label{eq_l}
\end{equation}
These relations are fits to a large set of cold stream simulations and are only approximate. 

We eliminate the common variable $M_8$ between Equations~\ref{eq_R} and \ref{eq_l} 
to find the gap creation rate as a function of gap size is,
\begin{equation}
R_\cup = 0.060 r^{0.44} \ell^{-1.16} \, {\rm kpc}^{-1} {\rm Gyr}^{-1},
\label{eq_rate}
\end{equation}
where the rate is the sum for all gaps larger than $\ell$, which is measured in kpc and the variable $r$ is scaled to 30 kpc.

For convenience, we convert Eq.~\ref{eq_rate} into variables appropriate to the observational quantities for the GD-1 stream.
A typical radius is approximately 15 kpc, so $r=0.5$ and  and we express gap sizes in terms of their measured 
angular sizes at 8.4 kpc, so $\ell = 0.14 \theta$, where $\theta$ is in degrees. 
Taking the rate to be constant over the 8.4 kpc stream segment and a mean lifetime of 4.6 Gyr gives,
\begin{equation}
N(>\theta) \simeq 15.8 t_{4.6}\,\theta^{-1.16},
\label{eq_nw}
\end{equation}
which is displayed as the solid line in Figure~\ref{fig_ngap}. 
The relation for a  2.3 Gyr mean stream age is shown as the dashed line. 

\begin{figure}
\begin{center}
\includegraphics[scale=0.8]{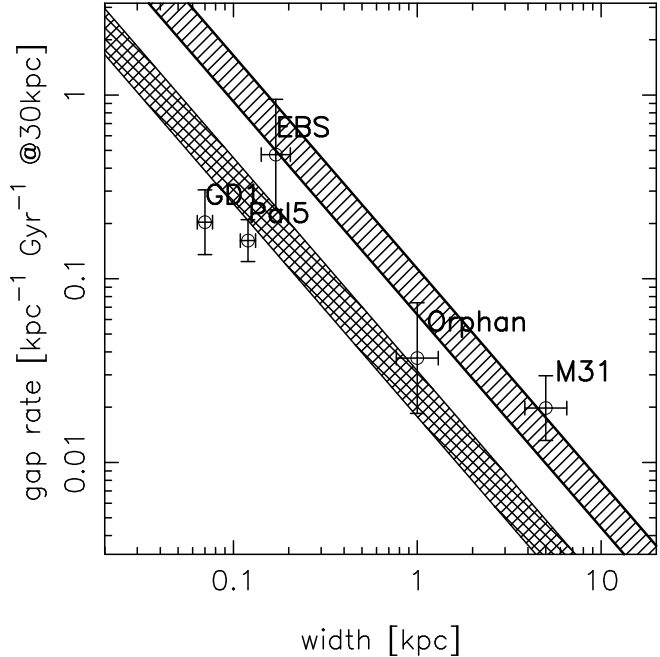}
\end{center}
\caption{The relation between the estimated gap creation rate per unit stream length and the width of the stream. The points are for the streams discussed in Table~\ref{tab_gw}, although only Pal~5 and GD-1 have been analyzed with our gap-filtering methods. 
The hatched region is the theoretical relation for cold streams developed from \citet{CGH:12}.  
The cross-hatched region assumes that the minimum visible gap is 3 times the stream width. For clarity the hatched regions are half the width used in \citet{CGH:12} and underestimate the uncertainty in the theoretical relation.}
\label{fig_gapwidth}
\end{figure}

The predicted relations in Figure~\ref{fig_ngap} have about the right slope and amplitude for gaps larger than 2\degr, 
but the number of gaps smaller than about 2\degr\ falls well below the prediction. 
The relationship assumes that gaps as small as the width of the stream are visible, whereas our gap analysis finds far fewer. 
The warm stream simulations of \citet{YJH:11} chose parameters modeled on the somewhat wider Pal~5 stream, 
finding that sub-halos of mass $3\times10^5\msun$ produced gaps so small that they would not likely be visible 
at our signal-to-noise. 
Our own preliminary simulation studies of streams of finite width find that  gaps as narrow as the stream width are created 
if a sub-halo with a scale radius greater than the stream width passes within 1.5 times the width of the stream.
Figure~\ref{fig_gapsM} shows the gap profiles in a stream of width  70 pc, 
on a circular orbit at 30 kpc radius that encounters sub-halos of mass $3\times10^5$, $10^6$ and $3\times 10^6\msun$ at distances of 90 and 150 pc.  
These encounters are all for a single but reasonably representative orbit locally perpendicular to the stream at 1.2 times the circular velocity. 
At $3\times 10^5\msun$ the gaps have a depth of $\simeq 30$\%, which will not be visible in the 30\% noise level of the GD-1 stream. 
At $10^6\msun$ the gaps are more than 60\% deep at 150 pc and are slightly more than 30\% deep 
at 300 pc. Therefore these simulations support that idea that stream interactions effectively cut off at a mass of $10^6\msun$ and for that mass at a distance from the stream centerline of about 3 times the stream width. 

In young streams  there is likely to be a significant range in angular momentum contributing to the stream width \citep{EB:11}. Producing a gap in the stream requires much heavier sub-halos, since the shear in the stream blurs out the gap \citep{YJH:11}.
The angular momentum spread at any given stream position declines with time and will not be a problem at our estimated age of GD-1.

We conclude that for the GD-1 stream the minimum effective sub-halo mass  is $10^6\msun$
which has a scale radius of 0.11 kpc. 
However, depending on the orbit of the progenitor and the dynamics of how the stars are stripped away the stream material may have sufficient orbital shear that small gaps may be blurred out into invisibility. 
The calculation of the total number of gaps needs to take the distribution of gap widths into account, which we do with a simple empirical adjustment. 

\subsection{Total Gap Rates}

Table~\ref{tab_gw} includes this paper's GD-1 measurements along with the star stream data previously 
presented in \citet{CGH:12} with the Pal~5 measurements. 
A simple revision to the cold stream approximation is to raise the minimum visible gap width to be $f$ times the width of the stream, so that the minimum visible width $w$ becomes $f w$.

The value of $f$ inferred from the GD-1 stream is 3 (from a FWHM of 0.5\degr\ to an effective minimum gap size of about 1.5\degr) which reduces the coefficient of Equation~\ref{eq_rate} by a factor of 3.6. 
The value of $f$ likely depends on the signal to noise in the stream and the residual angular momentum present along the
stream, which smooths out gaps, but the $f=3$ value should at least be representative.
The value of $f=3$ corresponds to increasing the minimum mass sub-halo that creates a visible gap
upwards by $3^{3.22}$, a factor of 35  increase,
which demonstrates that the inferred sub-halo masses 
are very sensitive to both the gap measurements and the dynamical details of the process.  

The adjusted rate relation is displayed in Figure~\ref{fig_gapwidth} as the cross-hatched region, 
along with the original cold-stream relation as the hatched region.  Note that only Pal~5 and GD-1 have had gaps counted with our gap-filtering approach. The other streams used simple counting estimates.
The agreement with 99\% confidence total counts of the Pal~5 and GD-1 streams  is quite good.

\begin{deluxetable}{rr rrr rrr}
\tablecolumns{8}
\tablewidth{0pc}
\tablecaption{Observed Stream Gap Statistics}
\tablehead{
\colhead{Stream} & \colhead{Gaps}   & \colhead{Length} & \colhead{Width}   & \colhead{Age/2} & \colhead{$R_{GC}$}& \colhead{$n/n_0$} & \colhead{$\mathcal R_\cup$} \\
 & \# & kpc & kpc & Gyr  &kpc & &  kpc$^{-1}$  \\
 & &  &  &   & & & Gyr$^{-1}$  \\
}
 \startdata 
M31 		& 12 	& 200 	& 5 			& 5			& 100		&6	 	& 0.012\\
Pal~5		& 6		& 8.1	& 0.12		& 3.5		& 19		&22		& 0.17\\
EBS			& 8		& 4.7	& 0.17		& 3.5		& 15 		&24 	&0.49	\\
Orphan 		& 2 		& 30 	&1.0		& 1.8		& 30		&17 	&0.037\\
GD-1		& 8		& 8.4	& 0.070		& 4.6		& 15		&24		&0.21 \\
\enddata
\label{tab_gw}
\end{deluxetable}
%65degrees at 8kpc

\subsection{A Simple Calculation of Sub-halo numbers}

A simple model independent volumetric density for the objects creating the gaps, $n$, can be calculated 
from the basic scaling relationship for stream crossings, $R_\cup \simeq \pi nfwv_{rel}$, 
where $v_{rel}$ is the typical stream crossing velocity and $w$ the stream width.
Taking a representative velocity to be the GD-1 orbit estimated circular velocity  of  220 \kms\ \citep{KRH:10},
then with our gap creation rate of 0.21 kpc$^{-1}$ Gyr$^{-1}$, gives a
density of gap-creating objects of 0.0015 kpc$^{-3}$ for $f=3$. 
The total number of objects over the 30 kpc extent of the GD-1 orbit is then about 90, 
where we have used the Aquarius density profile to compute numbers. 
This is quite close to 110 sub-halos we estimate in the same volume on the basis of
our $10^6\msun$ mass cutoff  in the more elaborate calculations given above.
These masses and numbers are in reasonable accord given the uncertainties and the strong dependence on the mass on the input numbers.  

\section{CONCLUSIONS}

GD-1 is an exceptional star stream for the study of the statistics of stream gaps. The segment we examine has a width of 70 pc and a length of 8.4 kpc, hence a length to width ratio above one hundred.
 GD-1's signal to noise  is relatively high at about 3 per longitudinal bin of 0.1\degr\ (0.014 kpc) 
 which enables a first measurement of the distribution of gaps sizes. 
 Our study of the performance of our  gap search procedure finds that it generates 44\% false positives even at 
 our 99\% confidence level. 
 This bias is taken into account in our gap counts, leading to $8\pm3$ gaps at 99\% confidence.
 We find that the larger gaps are in good agreement with model prediction. However,
gaps less than 3 times the stream width fall well below the cold stream relation. 
We therefore set the minimum visible gap size to 3 times the width of the stream, which is supported by some
preliminary warm stream simulations. 
The revised total gap creation rate is in good agreement with the GD-1 data, and data from other streams. 
The total number of stream gaps directly indicates a population of
about 100 sub-halos within the 30 kpc orbital extent of GD-1, which extrapolates to more than 15,000 within
the 433 kpc normalizing radius of the Milky Way halo model. 
These numbers are well in excess of any known population and
 are in good agreement, within the random errors and systematic error of stream age, 
 with the distribution and total numbers of predicted LCDM dark matter sub-halos with $M>10^6\msun$. 
 
 The idea of dark matter sub-halo stream crossings leads to a consistent picture of gaps in streams but it is premature to declare that the evidence is conclusive. 
At this stage there seems little doubt that gaps in streams are real and not readily explained in a smooth halo model.
There are several steps to improve the results. First the modeling is relatively straightforward to improve using 
 realistic progenitors on observed orbits within a full n-body approach. 
 Secondly, as impressive as the SDSS data is for stream measurements, upcoming data will go much deeper which will improve the signal-to-noise and discover new streams at larger galactocentric distances. 
 The streams are expected to show small offsets from the unperturbed path, comparable to the stream width, near gaps, which should become detectable.  
 And, as high precision wide field astrometric and kinematic data becomes available it will immediately help separate the stream from unrelated field stars and open up new opportunities to undertake kinematic modeling of streams and their gaps.

\acknowledgements

This research was supported by CIFAR and NSERC Canada. 
Thoughtful comments from an anonymous referee lead to improvements in content and presentation.

The SDSS is managed by the Astrophysical Research Consortium for the Participating Institutions. The Participating Institutions are the American Museum of Natural History, Astrophysical Institute Potsdam, University of Basel, University of Cambridge, Case Western Reserve University, University of Chicago, Drexel University, Fermilab, the Institute for Advanced Study, the Japan Participation Group, Johns Hopkins University, the Joint Institute for Nuclear Astrophysics, the Kavli Institute for Particle Astrophysics and Cosmology, the Korean Scientist Group, the Chinese Academy of Sciences (LAMOST), Los Alamos National Laboratory, the Max-Planck-Institute for Astronomy (MPIA), the Max-Planck-Institute for Astrophysics (MPA), New Mexico State University, Ohio State University, University of Pittsburgh, University of Portsmouth, Princeton University, the United States Naval Observatory, and the University of Washington.

{\it Facility:} \facility{Sloan Digital Sky Survey}

\end{document}